\documentclass{aa}  
\usepackage{graphicx}
\usepackage{txfonts}
\usepackage{float}
\usepackage{xcolor}
\usepackage{natbib,twoopt}
\usepackage[breaklinks=true]{hyperref} 
\bibpunct{(}{)}{;}{a}{}{,} 
\makeatletter
\newcommandtwoopt{\citeads}[3][][]{\href{http://adsabs.harvard.edu/abs/#3}%
	{\def\hyper@linkstart##1##2{}%
		\let\hyper@linkend\@empty\citealp[#1][#2]{#3}}}
\newcommandtwoopt{\citepads}[3][][]{\href{http://adsabs.harvard.edu/abs/#3}%
	{\def\hyper@linkstart##1##2{}%
		\let\hyper@linkend\@empty\citep[#1][#2]{#3}}}
\newcommandtwoopt{\citetads}[3][][]{\href{http://adsabs.harvard.edu/abs/#3}%
	{\def\hyper@linkstart##1##2{}%
		\let\hyper@linkend\@empty\citet[#1][#2]{#3}}}
\newcommandtwoopt{\citeyearads}[3][][]%
{\href{http://adsabs.harvard.edu/abs/#3}
	{\def\hyper@linkstart##1##2{}%
		\let\hyper@linkend\@empty\citeyear[#1][#2]{#3}}}
\makeatother

\usepackage[normalem]{ulem}
\usepackage{soul}

\begin{document}

   \title{The role of the thermal properties of electrons on the dispersion properties of Alfvén waves in space plasmas}


\author{N. Villarroel-Sep\'{u}lveda
	\inst{1,2}
	\and
	P. S. Moya \inst{1}
	\and 
	R. A. L\'{o}pez \inst{3} 
	\and
	D. Verscharen \inst{2}  
}
\institute{Departmento de F\'{\i}sica, Facultad de Ciencias,
	Universidad de Chile, Las Palmeras 3425, 7800003, Ñuñoa, Santiago, Chile\\
	\email{nicolas.villarroel@ug.uchile.cl, pablo.moya@uchile.cl}
	\and
	Mullard Space Science Laboratory, University College London, Dorking, RH5 6NT, UK
	\and 
	Research Center in the intersection of Plasma Physics, Matter, and Complexity ($P^2 mc$), Comisi\'on Chilena de Energ\'{\i}a Nuclear, Casilla 188-D, Santiago, Chile
}

 \abstract 
   {The transition from left-hand to right-hand polarized Alfvén waves depends on the wavenumber, the ratio of kinetic to magnetic pressure $\beta$, temperature anisotropy, and ion composition of the plasma. Along with the temperature anisotropy, the electron-to-proton temperature ratio $T_e/T_p$ is of great relevance for the characterization of the thermal properties of a plasma. This ratio varies significantly between different space plasma environments. Thus, studying how variations on this ratio affect the polarisation properties of electromagnetic waves becomes highly relevant for our understanding of the dynamics of space plasmas.}
   {We present an extensive study on the effect of the thermal properties of electrons on the behaviour and characteristics of Alfvénic waves in fully kinetic linear theory, as well as on the transition from electromagnetic ion-cyclotron (EMIC) to kinetic Alfvén waves (KAW).}
   {We solve the fully kinetic dispersion relation for oblique electromagnetic waves of the Alfvén branch in a homogenous Maxwellian electron-proton plasma. We quantify the effect of the thermal properties of electrons by varying the electron-to-proton temperature ratio for different configurations of the propagation angle, $\beta_p=8\pi nkT_p/B^2$, and wavenumber.}
   {We show that the temperature ratio $T_e/T_p$ has strong and non-trivial effects on the polarisation of the Alfvénic modes, especially at kinetic scales ($k_{\perp}\rho_L>1$, where $k_{\perp}=k \sin\theta$ and $\rho_L= c_s/\Omega_p$, with $c_s$ the plasma sound speed and $\Omega_p$ the proton’s	gyrofrequency) and $\beta_e+\beta_p>0.5$. We conclude that electron inertia plays an important role in the kinetic scale physics of the KAW in the warm plasma regime, and thus cannot be excluded in hybrid models for computer simulations.} 

\maketitle
\section{Introduction}
Since the first half of the XX century, plasma waves have become fundamental for the study and comprehension of the dynamics of space and astrophysical plasmas. Many of these physical environments, such as the solar wind, planetary magnetospheres, and outer heliosphere \citep{Livadiotis_2017}, as well as supernova remnants \citep{Barnes_Scargle_1973, Kato_2007}, are weakly collisional, meaning that the long-range forces due to the electromagnetic fields dominate over the short-range forces from particle-particle interactions \citep{Landau_Lifshitz_1981}. In this type of plasma, where the effect of particle collisions may be neglected, wave-particle interactions play a major role in the regulation of the large-scale global behavior of the system. Fluctuations in the electromagnetic fields extend over a broad range of spatial and temporal scales, with energy flowing from the larger MHD scales to the smaller kinetic scales, where the energy is dissipated as the plasma is heated. Mechanisms from linear (or quasi-linear) kinetic theory, like plasma instabilities and Landau damping, can effectively exchange energy between the particles and the oscillating electromagnetic fields, shaping the velocity distribution of the plasma \citep{Gary_1992,Livadiotis_2017, Verscharen_2019, Roberts2022}, and have been therefore proposed as candidates for the collisionless dissipation of the energy cascade, although non-linear effects like turbulent decay may also play an important role \citep{Matthaeus_2020}. Thus, the study of plasma wave modes from kinetic theory and their dispersion properties acquire a fundamental role in our understanding of the small-scale physics of space plasmas. 

One of the candidate wave modes that forms part of the turbulent spectrum at sub-proton scales is the kinetic Alfvén wave (KAW), as the power spectrum and characteristics of the electromagnetic fluctuations exhibit signature properties of this mode both in the solar wind \citep{Salem_2012} and terrestrial magnetosphere \citep{Roberts2022}, and can be therefore be expected do the same in other magnetized astrophysical plasma environments \citep{Alfven_1987}. The KAW is the small-scale extension of the MHD shear-Alfvén when kinetic effects acquire relevance and the wave develops fluctuations in the electrostatic potential and the field-parallel component of the electric field. In low $\beta$ plasmas ($\beta<m_e/m_p$, where $\beta=8\pi nk_BT/B^2$ in a one-temperature isotropic plasma and $m_e$, $m_p$ are the mass of electrons and protons, respectively), the KAW (often referred to as inertial Alfvén wave, in this particular case,) appears when the perpendicular wavelength becomes of the order of the electron's inertial length, and the wave is driven by electron inertial effects \citep{goertz_boswell_1979, lysak_lotko_1996, Nandal2016, barik_singh_lakhina_2019c, barik_singh_lakhina_2020}. In higher beta plasmas ($\beta>m_e/m_p$), thermal effects dominate, and the wave appears when the perpendicular wavelength becomes comparable to the proton's (or ion's) Larmor radius \citep{Hasegawa1977, Hollweg1999}. The latter is the case that corresponds to the $\beta$ regime that is considered for this article. A detailed discussion and even an illustrative cartoon regarding the difference between the inertial and kinetic limits of the Alfvén wave are provided by \cite{Chen_Boldyrev_2017}, \cite{barik_singh_lakhina_2021}, and \cite{Barik_2023}. 

KAWs have been widely observed in a varied range of space plasma environments; spacecraft measurements have provided direct evidence of this mode propagating in Earth's magnetopause \citep{Chaston2005, Chaston2007, Yao2011}, plasma sheet \citep{Wygant2002, chaston_bonnell_2012, Stawarz2017, Zhang2022}, magnetosheath \citep{Travnicek_2022}, and inner magnetosphere \citep{Chaston2014, Chaston2015b,Malaspina2015, Moya2015, saikin_2015, Tian2022}, as well as in the solar wind \citep{he_tu_marsch_yao_2011, Perschke2013, Salem_2012, he_wang_tu_marsch_zong_2015, Huang_2020}. Furthermore, evidence of these waves through detection of particle energization along magnetic field lines has been observed in the heliosphere \citep{wu_yang_2006, artemyev_zimovets_rankin_2016} as well as in Earth's \citep{Artemyev2015, Tian2022, Zhang2022} and Jupiter's magnetospheres \citep{Gershman_2019, Sulaiman_2020, Damiano_2023}. The large number of observations of KAWs in different astrophysical environments and their theoretical prediction in many others prove them ubiquitous in space plasma environments, making their study essential for understanding the underlying physics behind many of the phenomena observed in this kind of systems, such as collisionless plasma heating \citep{Gershman__2017}, electron energization \citep{Zhou_Liu_Loureiro_2023}, and magnetic reconnection \citep{Boldyrev2019}.

One of the signature properties of KAWs is their right-handed polarisation; in the plane transverse to the background magnetic fields, the perturbed electromagnetic fields rotate in the sense of the electron's Larmor gyration \citep{gary_1986}. This makes this mode resonant with electrons, which leads to wave-particle interactions at sub-proton scales. As the propagation angle decreases, the transverse polarisation becomes linear and shifts to left-handed as the wave transitions to the quasi-parallel electromagnetic ion-cyclotron (EMIC) wave. The EMIC mode is resonant with the Larmor gyration of positive ions and is of great relevance in the plasma dynamics of Earth's magnetosphere \citep{Usanova_Mann_2016, Usanova_2021}. The transition angle between the KAW and EMIC modes has been shown to depend heavily on the wavenumber and plasma beta \citep{gary_1986, Hollweg1999, hunana_2013}, as well as on the temperature anisotropy \citep{MOYA2021} and ion composition of the plasma \citep{Moya2015, MOYA2021, Moya2022, Villarroel_2023}. These studies, however, consider fixed temperature ratios between electrons and ions. The ion-to-electron temperature ratio $T_i/T_e$ varies considerably on average between different astrophysical environments, and it also may vary in time between a large range of values in a single system \citep{Wilson_2018}. In Earth's plasma sheet, the most frequently observed values are $3\leq T_i/T_e \leq 5$ \citep{Grigorenko_2016}; in Earth's magnetotail, $T_i/T_e\sim 3$, while $5\leq T_i/T_e \leq 7$ can be observed for low electron temperatures \citep{Artemyev_2011}; other meassurements in the magnetosheath and plasma sheet have found that $6\leq T_i/T_e \leq 10$ when the plasma is cool, and $2\leq T_i/T_e \leq 5$ when the plasma is warm \citep{Wang2012}; in the solar wind near 1 au, most frequent observations show $0.28\leq T_i/T_e\leq 0.47$ with an average of $T_i/T_e\sim0.61$, with important variations during coronal mass ejections \citep{Wilson_2018, Mizuno}; in the solar corona, $T_i/T_e \sim 3.5$ has been observed \citep{Boldyrev_2020}; observations in Balmer-dominated supernova remnants show $3\leq T_i/T_e\leq10$ \citep{van_Adelsberg_2008}, while in astrophysical shocks \citep{Raymond_2023} and low-luminosity accretion disks \citep{Shapiro_Lightman_Eardley_1976,Quataert_Narayan_1999,Mizuno} the ion temperature can surpass that of electrons by several orders of magnitude. Thermal and inertial effects are known to play important roles in the kinetic dispersion properties of plasma waves, and the temperature ratio between electrons and ions is thus expected to affect the transition from the EMIC to the KAW mode. 

 The role of electron properties from a fully kinetic theory has been studied with a focus on the fire-hose instability \citep{Maneva_2016, Lopez_2022}. \cite{Howes_2009} studied the limitations of Hall MHD to Vlasov-Maxwell theory, comparing the dispersion relations of whistler and Alfvén waves obtained from both theories. He found that discrepancies between both theories depend on the wavenumber, propagation angle, $T_i/T_e$, and $\beta_p$. Previous studies on the effect of electron properties on the dispersion of KAWs have used fluid approximations \citep{Jana_2017} or taken the limit of near-perpendicular propagation in kinetic theory \citep{Quataert_1998,Agarwal_2011, Tong_2015, Zhao_2015}. A very complete study regarding  In this study, we present an extensive investigation of the effect of the ion-to-electron temperature ratio on the dispersive properties of KAWs from a full kinetic theory. This will allow us to understand the effect of electron properties on the transition from the EMIC to the KAW in astrophysical plasmas and develop insight into what kinetic scales and how electron inertia becomes relevant for the propagation of these waves.\\

\section{Linear theory for a weakly collisional plasma} \label{sec:style}

The particles of each of the species comprising the plasma are described statistically by phase-space distribution functions $f_s(\mathbf{r},\mathbf{v},t)$, where the sub-index $s$ represents a particular species.

These distribution functions satisfy Vlasov's equation for a weakly collisional plasma

\begin{align}
    \frac{\partial f_{s}}{\partial t}+\mathbf{v}\cdot\nabla_{\mathbf{r}}f_{s}+\frac{q_s}{m_s}[\mathbf{E}+\frac{\mathbf{v}}{c}\times\mathbf{B}]&=0, \label{Eq: Vlasov}
\end{align}

where $\nabla_{\mathbf{r}}$ is the usual spatial gradient and $\nabla_{\mathbf{v}}$ is the gradient in velocity space, and $q_s$ is the electric charge of the $s$-th species.
The system is coupled with Maxwell's equation through the charge and current densities:

\begin{align}
    \rho(\mathbf{r},t)&=\sum_{s}q_s\int f_s(\mathbf{r},\mathbf{v},t)d^3v, \\
    \mathbf{j}(\mathbf{r},t)&=\sum_{s}q_s\int \mathbf{v}f_{s}(\mathbf{r},\mathbf{v},t)d^3v.
\end{align}

We consider linearized quantities, such that the system is in equilibrium at zeroth order and perturbations, preceded by the letter $\delta$, vary rapidly in time:

\begin{align}
    f_{s}(\mathbf{r},\mathbf{v},t)&=f_{0s}(\mathbf{v})+\delta f_{s}(\mathbf{r},\mathbf{v},t),\\
    \mathbf{E}(\mathbf{r},t)&=\delta\mathbf{E}(\mathbf{r},t),\\
    \mathbf{B}(\mathbf{r},t)&=\mathbf{B}_0 + \delta\mathbf{B}(\mathbf{r},t).
\end{align}
With the stable equilibrium solution of Vlasov's equation given by a Maxwellian velocity distribution function given by
\begin{equation}\label{Eq1}
    f_{s}(v_\parallel,v_\perp)=\frac{n_{s}}{\pi^{3/2}\alpha_{s}^3 }\exp \left\{-\frac{v^2}{\alpha_{s}^2}\right\}, 
\end{equation}
for each particle species, where $v^2=v_{\perp}^2+v_{\parallel}^2$ with $v_{\parallel, \perp}$ indicating the kinetic field-aligned and transverse velocity of the particles, respectively. The thermal speed of each species is given by $\alpha_{s}=\sqrt{2k_BT_{s}/m_s}$, where $m_s$ and $T_s$ are the mass and temperature of the corresponding species, while $n_s$ is the particle density. We are interested in studying the oblique propagation of electromagnetic waves. We, therefore, set the propagation direction, without loss of generality, as $\mathbf{k}=k_\perp\hat{\mathbf{x}}+k_\parallel\hat{\mathbf{z}} = k\sin\theta\hat{\mathbf{x}}+k\cos\theta\hat{\mathbf{z}}$, where the propagation or wave-normal angle $\theta$ is the angle between the wave-vector and the background magnetic field $\mathbf{B}_0=B_0\hat{\mathbf{z}}$. 

Following Landau's approach \citep{Landau_1965}, we use Laplace transforms for time and Fourier transforms for the spatial coordinates of the linearized quantities, and by treating (\ref{Eq: Vlasov}) in reciprocal space, we obtain, at first order, the dispersion relation

\begin{equation}
 \mathbf{D}_k\cdot\delta\mathbf{E}_k=0, \label{disp}
\end{equation}
where 

\begin{align}
    \mathbf{D}_k&=\begin{pmatrix}
    D_{xx}&iD_{xy}&D_{xz}\\
    -iD_{xy}&D_{yy}&iD_{yz}\\
    D_{xz}&-iD_{yz}&D_{zz}
    \end{pmatrix}
\end{align}
is the full dispersion tensor for oblique waves (see \citep{Villarroel_2023} for further detail), and the eigenvectors $\delta\mathbf{E}_k$ represent the electric field perturbations. The Laplace transform on the time-coordinate implies that the solutions will be damped oscillations, and we will refer to the real part of the complex frequency as $\omega$ and to the imaginary part as $\gamma$. With the information of the dispersion tensor, we can compute the polarisation of the transverse component of the waves as taken with respect to the background magnetic field, as defined by \citet{Stix1992}, using


\begin{equation}
   i\frac{\delta E_{kx}}{\delta E_{ky}}=\frac{D_{xy}D_{zz}-D_{xz}D_{yz}}{D_{xx}D_{zz}-D_{xz}^2}\textnormal{sign}(\omega). \label{polStix}
\end{equation} 
By convention, a right(left)-handed polarisation is defined by a timewise rotation of the fields according to the right(left)-hand rule around the magnetic field $\mathbf{B}_0$, such that field-perpendicular electric fields in a right-hand polarized wave gyrate in the sense of the Larmor motion of electron, while in a left-hand polarized wave the fields rotate in the sense of the Larmor gyration of positive ions.\\

The MHD Alfvén wave is linearly polarized, with its electric perturbations appearing only in the $\hat{\mathbf{x}}$ direction, such that the transverse electric polarisation \eqref{polStix} is infinite in $k\to0$, and becomes finite as $E_y$ becomes different than $0$ when kinetic effects become relevant, indicating that the wave becomes elliptically polarized. In general, the complex nature of the fields implies that they possess both linear and elliptical polarisation, with $\textnormal{Re}[i\delta E_{kx}/\delta E_{ky}]$ being a measure of the mode's elliptical polarisation and $\textnormal{Im}[i\delta E_{kx}/\delta E_{ky}]$ representing the linear polarisation of the wave. We are particularly interested in the elliptical polarisation of the waves, so the quantity to be analyzed will be the real part of \eqref{polStix}. We note, however, that the polarisation is completely linear whenever $\textnormal{Re}[i\delta E_{kx}/\delta E_{ky}]=0$ or $\textnormal{Re}[i\delta E_{kx}/\delta E_{ky}]=\pm\infty$. Circular polarisation occurs when $\textnormal{Re}[i\delta E_{kx}/\delta E_{ky}]=\pm1$; where right-handed is obtainded when the RHS equals $+1$, and left-handed polarisation occurs for $-1$.

Other polarisation ratios can be obtained from (\ref{disp}), such as 

\begin{align}
    i\frac{\delta E_{kz}}{\delta E_{ky}}&=\frac{D_{xx}D_{yz}-D_{xz}D_{xy}}{D_{xx}D_{zz}-D_{xz}^2}\textnormal{sign}(\omega). \label{polz}
\end{align}

Replacing both (\ref{polStix}) and (\ref{polz}) in Faraday's law allows us to evaluate 

\begin{align}
    i\frac{\delta B_z}{\delta B_y}&=\frac{ik_\perp E_{ky}}{k_{\parallel}E_{kx}
    -k_{\perp}E_{kz}
    }\nonumber\\
    &=\frac{D_{xz}^2-D_{xx}D_{zz}}{\cot\theta(D_{xy}D_{zz}-D_{xz}D_{yz})+(D_{xz}D_{xy}-D_{xx}D_{yz})}\textnormal{sign}\big(\frac{\omega}{k}\big),\label{polmag0}
\end{align}

whose characteristic behavior for the KAW has been thoroughly analyzed in two-fluid theory by \cite{Hollweg1999}, who obtained the first and second-order finite Larmor radius corrections to this quantity, which is zero in MHD:

\begin{subequations}
\begin{align}
    i\frac{\delta B_z}{\delta B_y}&\approx-i\frac{k_{\perp}v_s^2}{v_A\Omega_{p}}\textnormal{sign}(\omega/k_z), \label{polmag1}\\
      i\frac{\delta B_z}{\delta B_y}&\approx -i\frac{k_{\perp}v_s^2}{v_A\Omega_{p}(1+\frac{k_\perp^2v_{sp}^2}{\Omega_p^2})}\textnormal{sign}(\omega/k_z),, \label{polmag2}
\end{align}
\end{subequations}
where $v_s=[(\gamma_p T_p+\gamma_eT_e)/m_p]^{1/2}$ is the plasma sound speed, with $\gamma_s$ the polytropic index of species $s$, $\Omega_s=\frac{q_sB}{m_s}$ is the gyrofrequency of the $s-$th plasma species, with $q_s$ its charge, and $v_{sp}=(\gamma_pT_p/m_p)^{1/2}$.

The dispersion relation of the KAW can be obtained directly by assuming $k_{\perp} \gg k_{\parallel}$, $m_{e}/m_{p}\ll\beta\ll1$, and $|\omega|\gg|\gamma|$ (see \cite{lakhina_1990}, \cite{lakhina_2008}, \cite{barik_singh_lakhina_2019a}, \cite{barik_singh_lakhina_2019b}, for example). Since we are interested in studying the transition of left-hand polarized Alfvén waves to their right-hand polarized counterparts, which depends on both the propagation angle and the plasma temperature, and because the waves can become strongly damped for smaller angles, we must relax the aforementioned approximations and solve the full dispersion relation of oblique Alfvénic waves using the DIS-K solver\footnote{The full code is publicly available and can be found at \url{https://github.com/ralopezh/dis-k}.} \citep{Lopez_2021,Lopez_disk} in the limit of isotropic Maxwellian distributions as given by Eq.~(\ref{Eq1}). We have checked our results by comparing them with the NHDS code \citep{Verscharen_2018}. Because no approximations are taken regarding the frequency of the waves, these solutions are valid even in the strong-damping wavenumber domain ($|\gamma(k)|\geq|\omega(k)|$, where $\gamma$ and $\omega$ represent the imaginary and real parts of the complex frequency, respectively). \\
Under Alfvén mode, we will refer to the continuous extension of the MHD Alfvén wave to the large-wavenumber domain, which, we will identify and differentiate from other solutions that possess right-handed polarisation at near-perpendicular angles by analyzing the phase velocity of the waves in the MHD limit (see Appendix B). It is worth noting that this extension may differ considerably from the cold plasma solutions because of the effects of finite temperature, especially for propagation angles that depart from near-parallel or near-perpendicular propagation. In this context, it is not unusual that the real frequency of the continuous kinetic extended MHD Alfvén wave may surpass the proton gyrofrequency at relatively large wavenumbers and has then been referred to as the Alfvén/whistler solution in two-fluid theory \citep{Sahraoui_2012}, although a similar effect can happen due to mixing between wave modes in kinetic theory \citep{Krauss_1994}. This phenomenon is very complex and highly dependent on the temperature of each of the plasma species \citep{isenberg_1984} and, although interesting by itself, its in-depth analysis and discussion lie beyond the scope of this article.\\
For a fixed value of the beta parameter, we obtain the real and imaginary parts of the wave's frequency normalized to the protons' cyclotron frequency $\Omega_{p}$, as well as their polarisation, all as a function of the wavenumber normalized to the inverse of the proton inertial length $\omega_{pp}/c$, with $\omega_{pp}$ the protons' plasma frequency, and $c$ the speed of light in vacuum.
\section{Results}\label{sec:results}
\begin{figure*}
    \centering
    \includegraphics[width=0.9\textwidth]{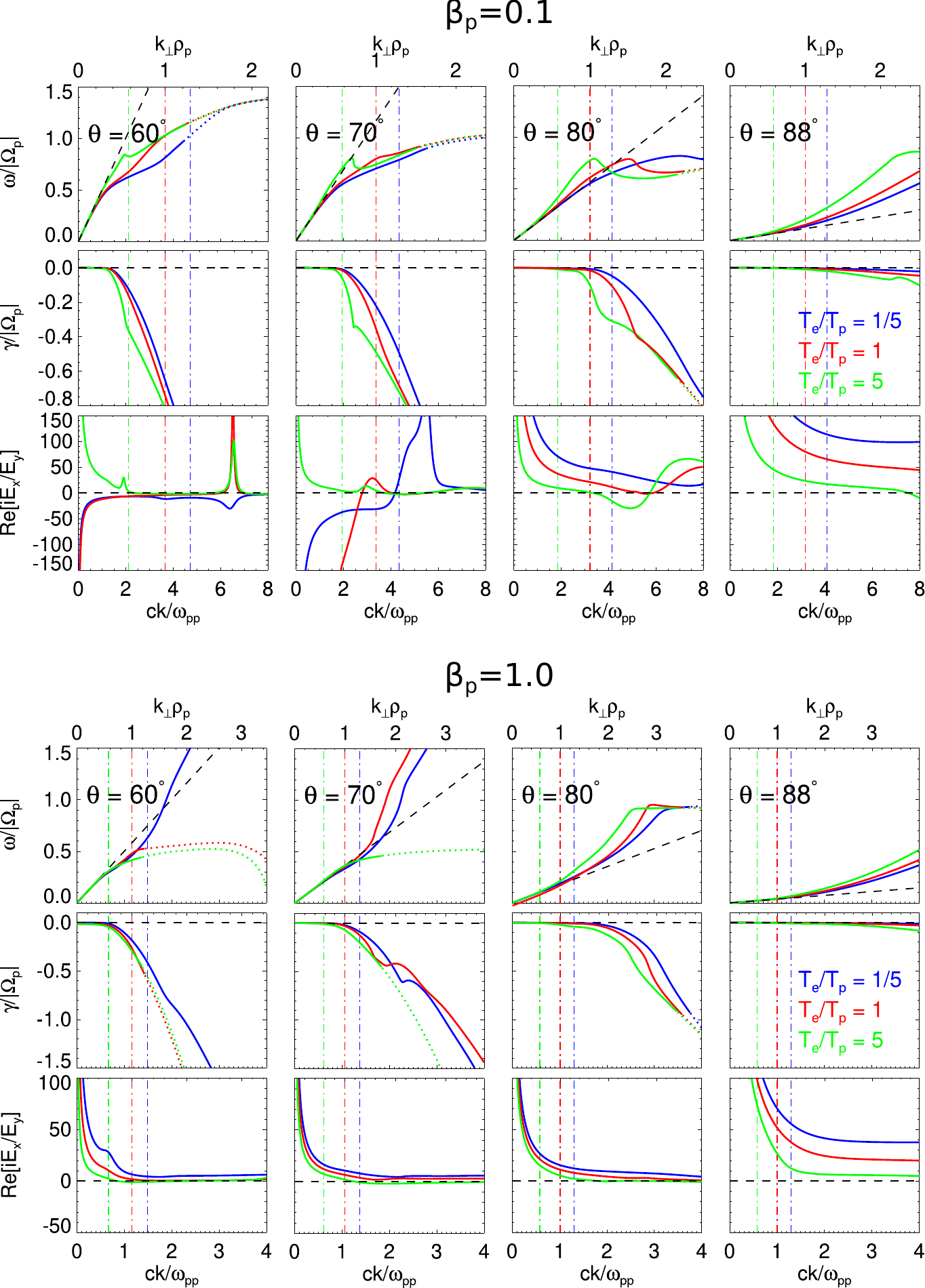}
    \caption{Real ($\omega$) and imaginary ($\gamma$) frequency of the Alfvén wave and its transverse polarisation for different electron-to-proton temperature ratios, with $\beta_p=0.1$ (top) and $\beta_p=1.0$ (bottom). The dispersion relation is plotted with continuous lines for  $|\gamma|<|\omega|$ and in dotted lines for $|\gamma|>|\omega|$, thus indicating that the mode is strongly damped. Dash-dotted lines indicate $k_{\perp}\rho=1$ for the different configurations, and the segmented line in the real frequency panels corresponds to the MHD dispersion relation of the Alfvén wave.}
    \label{fig:disp}
\end{figure*}

\begin{figure*}
    \sidecaption
    \includegraphics[width=12cm]{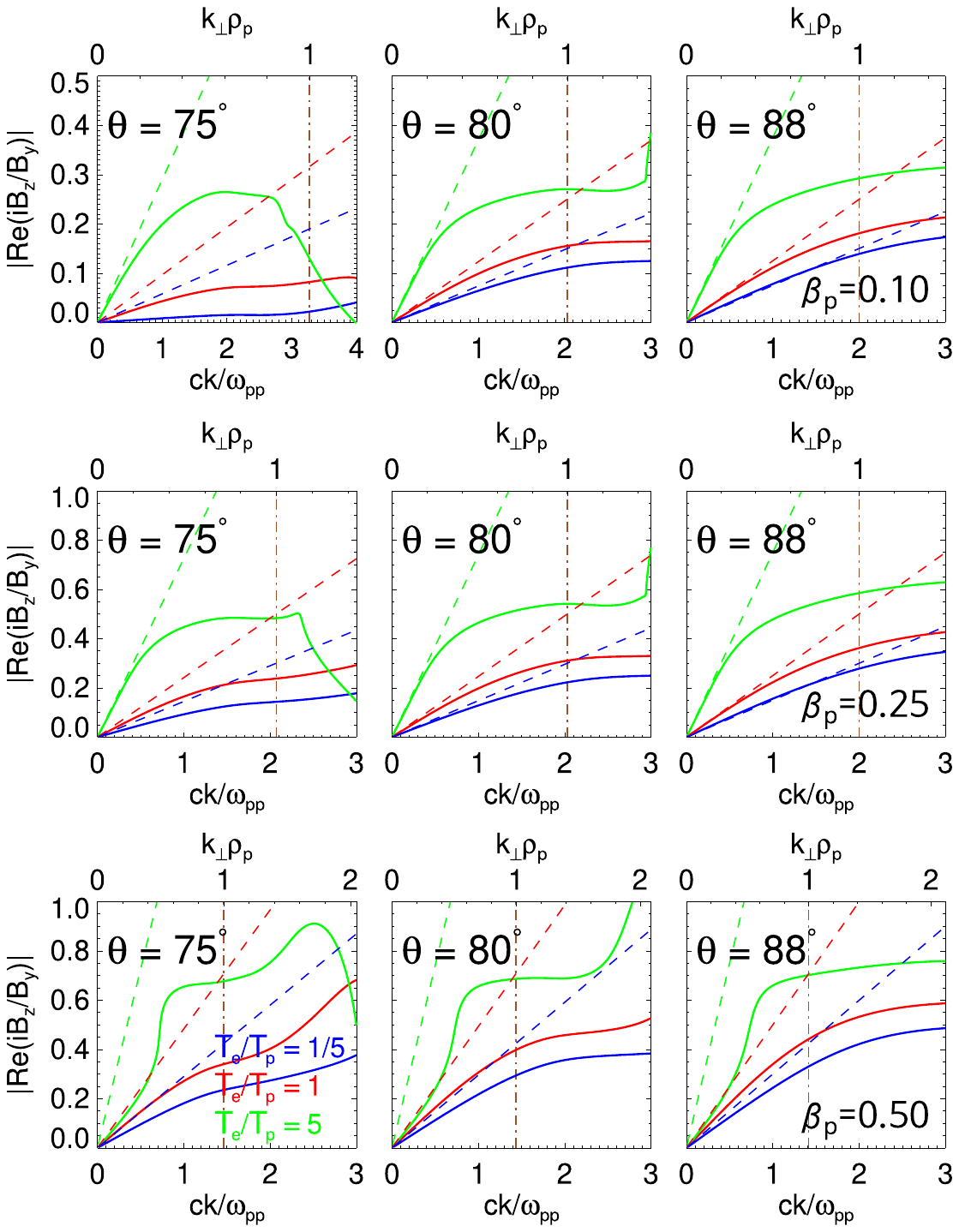}
    \caption{Magnitude of the real part of the magnetic ratio $iB_z/B_y$ for different values of $T_e/T_p$. Continuous lines are the results obtained from kinetic theory and indicate the rough predictions from the two-fluid theory given in \eqref{polmag1}.}
    \label{fig:pol_bzy}
\end{figure*}

Figure \ref{fig:disp} displays the real and imaginary frequency of the Alfvén mode, as well as its transverse electrical polarisation, for $\beta_p=0.1$ and $\beta_p=1.0$, as functions of the normalized wavenumber, with the different colours indicating different electron-to-proton temperature ratios. The solutions are plotted in different styles depending on the sign of $|\omega|-|\gamma|$ to make a distinction between weakly and strongly damped solutions, as strongly damped solutions will not effectively propagate and thus cannot represent realistic electromagnetic waves. \\
For all cases considered, as the propagation angle approaches perpendicular propagation, the waves acquire the typical characteristics of the KAW, with their polarisation becoming elliptically right-handed and their real frequency lying above MHD prediction. As can be seen in the Figure, this transition appears at different propagation angles in the different plasmas that are being considered. Both the proton beta $\beta_p$ and the electron-to-proton temperature ratio $T_e/T_p$ play an important role in this matter, as raising any of these quantities will effectively lower the transition angle, allowing right-hand polarized Alfvén waves to propagate at angles that depart from the near-perpendicular limit.  Vertical lines indicate $k_{\perp}\rho_{L}=1.0$ for the different curves, where $\rho_{L}=\frac{v_S}{\Omega_{p}}$ is referred to as the ion-acoustic Larmor radius \citep{Choi_Woo_Ryu_Lee_Yoon_2023}. This Larmor radius, which depends on both the electron and ion temperatures, can also be expressed as $\rho_L=\frac{c}{\omega_{pp}}\sqrt{\beta}$, with $\beta=(\frac{\beta_e+\beta_p}{2})$ being the ion-acoustic$\beta$, as defined in \citep{Choi_Woo_Ryu_Lee_Yoon_2023} and often utilized in two-fluid theory, with $\gamma_i=\gamma_e=1$. This choice of the polytropic indices is characteristic of an isothermic process and provides the correct dispersion relation for the magnetosonic waves in this plasma setting (see Appendix B of \cite{Villarroel_2023}), as well as the low-wavenumber limit of the magnetic polarisation (see Figure \ref{fig:pol_bzy}). 

Figure \ref{fig:pol_bzy} depicts in continuous colored lines the absolute value of this polarisation for different propagation angles, electron-to-proton temperature ratios, and $\beta_p$. Dashed lines indicate the first order two-fluid approximation of this quantity, given in \eqref{polmag1} with $\gamma_e=\gamma_p=1$. This expression provides an approximation for the behavior of this quantity in the quasi-perpendicular limit, although it deviates from the kinetic solutions at approximately the same wavenumbers as the first-order approximation and has, therefore, been omitted in the Figure in favor of readability. As is expected, for a fixed value of $\beta_p$, the wavenumber range in which this quantity and the prediction from fluid theory coincide becomes larger as the propagation angle approaches perpendicular propagation. This agreement in the small-wavenumber domain between the kinetic theory results and two-fluid theory predictions for the KAW is an indication that the displayed solution can be associated with the KAW most described in the literature. It is noteworthy that the agreement is not restricted to near-perpendicular propagation, as for $T_e>T_p$, the fluid prediction can be a good approximation even at $\theta=75^\circ$ for $\beta_p=0.25$. In general, if $\beta_p$ is sufficiently small, an increase in both $T_e/T_p$ and $\beta_p$ will allow the kinetic solution to coincide with the two-fluid KAW at a wider range of propagation angles. For large $\beta$, the fluid predictions exhibit completely different slopes than those of the results from kinetic theory, and thus cannot approximate the kinetic solutions even at near-perpendicular propagation.

\begin{figure*}
    \centering
    \includegraphics[width=0.9\textwidth]{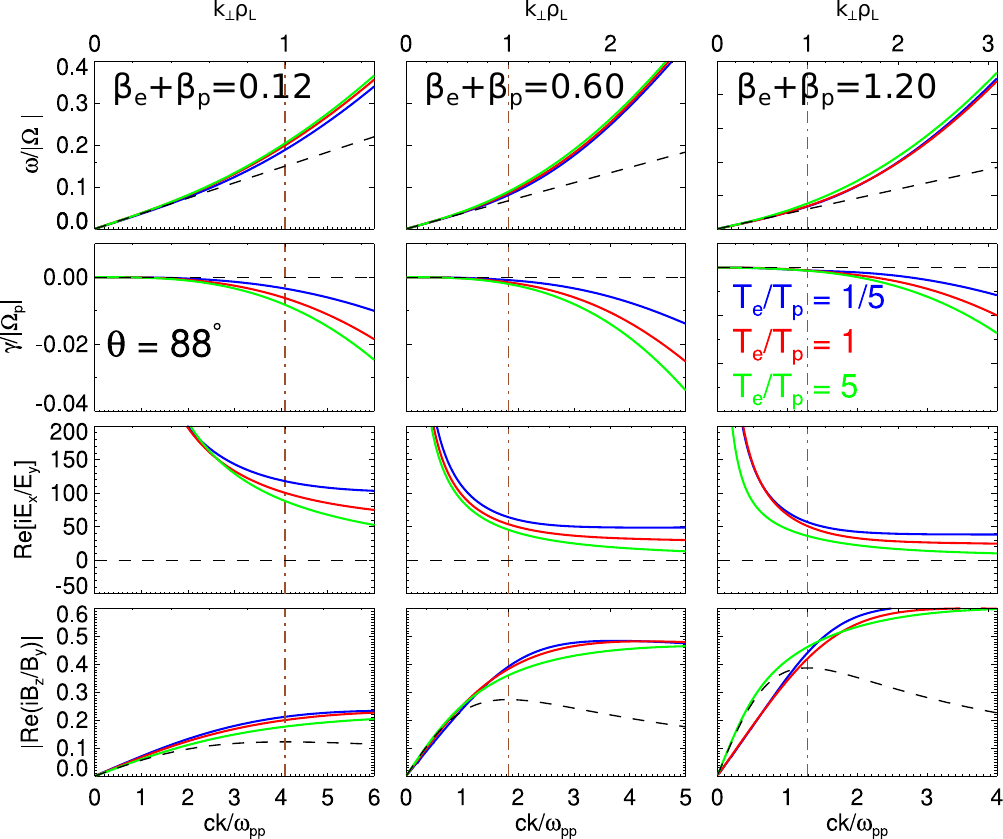}
    \caption{Real and imaginary frequency, transverse electrical polarisation, and magnetic polarisation of the Alfvén wave. The total beta of the plasma, $\beta =\beta_e+\beta_p$, is fixed at $0.12$, $0.60$, $1.20$, and for each value of $\beta$ different electron-to-proton temperature ratios are considered. Dash-dotted lines indicate $k_{\perp}\rho_{L}=1$ for the different values of $\beta_e+\beta_p$. The segmented line in the real frequency panels corresponds to the MHD dispersion relation of the Alfvén wave, in the two middle panels it indicates the line corresponding to zero, and in the bottom panel it represents the improved two-fluid approximation of this quantity given in equation \eqref{polmag2}.   }
    \label{fig:disp_fixedb}
\end{figure*}

Figure \ref{fig:disp_fixedb} depicts the real and imaginary frequency, transverse electrical polarisation (Eq. (\ref{polStix})) and magnetic polarisation (Eq. (\ref{polmag0})) for different values of $T_e/T_p$ and fixed values of $\beta_e+\beta_p$. As can be seen, deviations between the solutions considering different $T_e/T_p$  for a fixed total $\beta$ appear to be small even for $k_\perp \rho\geq 1$. Although differences in the transverse electrical polarisation reach considerable amounts, the behavior of these quantities remains qualitatively the same; the polarisation goes from linear at small wavenumbers to elliptical for larger wavenumbers, which can be seen in that $\textnormal{Re}[iE_x/E_y]$ decreases monotonically as $k$ increases, and plateaus for $k_\perp \rho \geq 1$.

\begin{figure*}
    \centering
    \includegraphics[width=0.75\textwidth]{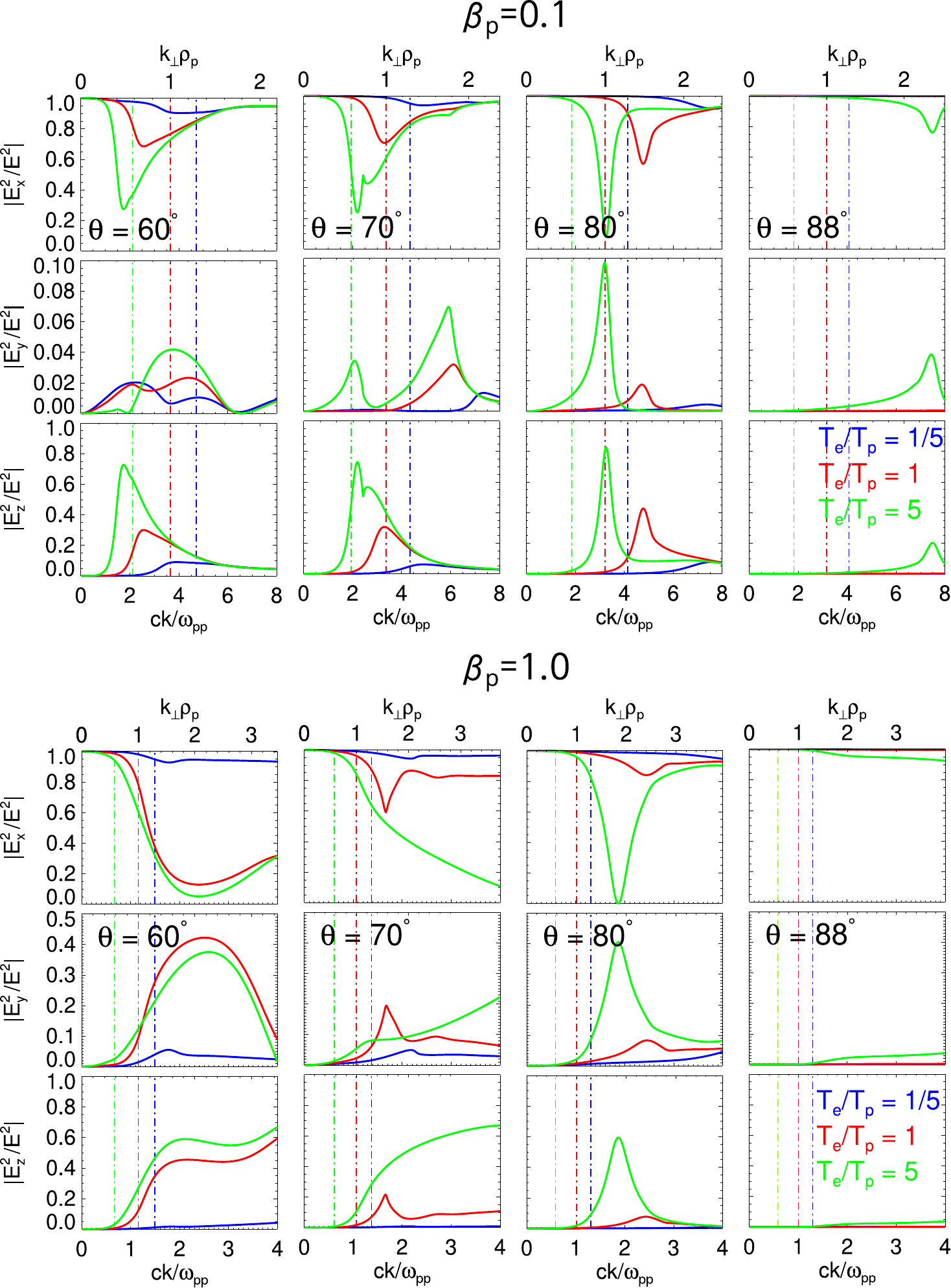}
    \caption{Amplitude of the perturbed electric field ratios for different values of $T_e/T_p$ and different propagation angles. Results are displayed for $\beta_p=0.1$ and $\beta_p=1.0$. Vertical dashed lines correspond to $k_{\perp}\rho_L=1.0$ for the corresponding electron-to-proton temperature ratio as indicated by the color code.}
    \label{fig:fields}
\end{figure*}

Figure \ref{fig:fields} shows the relative amplitude of the electric field components as functions of the wavenumber for different values of $T_e/T_p$, with $\beta_p=0.1$ and $\beta_p=1.0$. The linearized electric field appears to be sustained mostly by the field in the $x$ direction, which lies in the wave's propagation plane but is perpendicular to the background magnetic field. As the mode becomes dispersive, the field in the direction of the background magnetic field becomes excited at the expense of the perpendicular electric field. The way in which this occurs depends on $\beta$. For a smaller beta ($\beta_p=0.1$, in our case), the field in the $x$-direction decreases as its $z$-component increases until $|E_x^2|/|E^2|$ reaches a local minimum and $|E_z^2|/|E^2|$ a local maximum, and both quantities then plateau at large wave-numbers, with $|E_z^2|/|E^2|$ reaching a constant value different than $0$. The also perpendicular electric field in the $y$ direction is much smaller in magnitude but exhibits the same rise and decay as the one in the parallel direction. When $T_e>T_p$, the field's $z$-component may even become the dominant electric component for a considerable range of wave-numbers, although this does not take place in the near-perpendicular propagation limit. For a larger $\beta$, the asymptotic values of the amplitude ratios at large wavenumbers depend largely on the propagation angle and can differ significantly between the solutions with different $T_e/T_p$, which is not seen in the small $\beta$ case. Nevertheless, the core of the previous statement remains valid; the $E_x$ electric component recedes at larger wave-numbers as $E_z$ becomes large, although now $|E_y^2|/|E^2|$ can also achieve significant values which can be larger than those of the other amplitude ratios.  It is also worth noting that for the solution considering $\beta_p=0.1$ propagating at $\theta=70^\circ$, for $T_e/T_p=1$ and $T_e/T_p=5$ there are wavenumbers at which $|E_y^2|/|E^2|=0$. Since this is complex modulus, the equality implies that both the real and imaginary part of this field component must become zero. These wavenumbers coincide with local maxima in the corresponding transverse electrical polarisations depicted in Figure \ref{fig:disp}. Since these local maxima are finite, this implies that one of the components of $E_x$ must also become null. These zeroes in $E_y$ roughly coincides with the local maximum in $E_z$ and minimum in $E_x$, while for near-perpendicular propagation the minimum in $|E_x^2|/|E^2|$ coincides with maxima in the amplitude ratios $|E_z^2|/|E^2|$ and $|E_y^2|/|E^2|$. It is also worth mentioning that the nature of the excitation in the parallel electric field is different for different propagation angles and electron-to-proton temperature ratios. As $T_e/T_p$ becomes larger, the relative amplitudes of the $E_z$ and $E_y$ fields also become significantly larger. For larger wavenumbers, all cases approach the same constant value, implying the rise in the amplitude of the parallel electric field is likely a consequence of the inclusion of highly thermal electrons. Maxima in the parallel electric field appears to be displaced to larger wavenumbers as the wave approaches perpendicular propagation. Although for these propagation angles, the $E_z$ field does not become larger in magnitude than $E_x$, this drift towards small scales of the maxima of the relative amplitude of $E_z$ ensures the parallel electric field can acquire relatively high values in sub-proton scales, where wave-particle interactions are highly relevant.

\begin{figure*}[ht]
    \centering
    \includegraphics[width=0.75\textwidth]{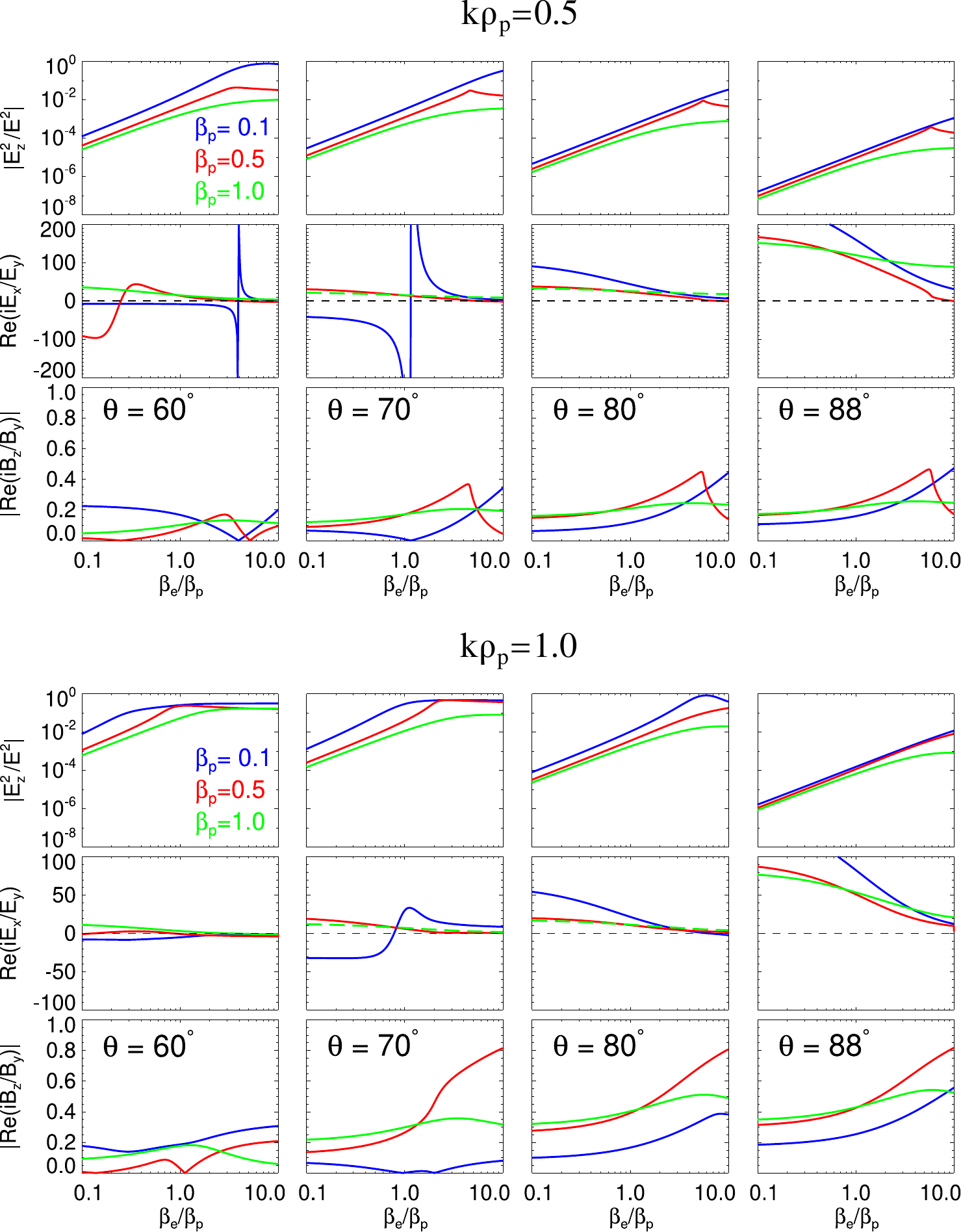}
    \caption{Amplitude of the electric field ratio $E_z/E_x$, real part of the transverse electric polarisation $iE_x/E_y$ and magnitude of the real part of the magnetic polarisation $iB_z/B_y$ for different values of $\beta_p$ and propagation angles. The results are displayed as functions of the electron-to-proton temperature ratio contained in $\beta_e/\beta_p$ for $k\rho_{p}=0.5$ and $k\rho_{p}=1.0$.}
    \label{fig:beta_pol}
\end{figure*}

Figure \ref{fig:beta_pol} displays the wave's parallel electric perturbations, its transverse electrical polarisation, and magnetic polarisation for different $\beta_p$ values as functions of the electron-to-proton temperature ratio for two fixed values of the wavenumber. The parallel perturbed electric field appears to be monotonically increasing with $T_e/T_p$ until it plateaus at a constant value, which is larger the smaller $\beta_p$ is. This plateau is also highly dependent on the propagation angle. For smaller propagation angles, the transverse electrical polarisation is susceptible to changes in sign. At $ck/\omega_{pp}=1.0$ with $\beta_p=0.1$ and $\beta_{p}=0.5$, an increase in $T_e/T_p$, which increases the total $\beta$, changes this polarisation from left-handed to right-handed, as it has been predicted by previous studies \citep{gary_1986}. In the latter case, however, since $|E_x|$ does not become zero in magnitude for any of the studied cases, the behavior suggests that at two points, $E_x/E_y$ and $B_z/B_y$ become purely real, indicating that the fields become linearly polarized but not zero.

\begin{figure*}
\sidecaption
\includegraphics[width=12cm]{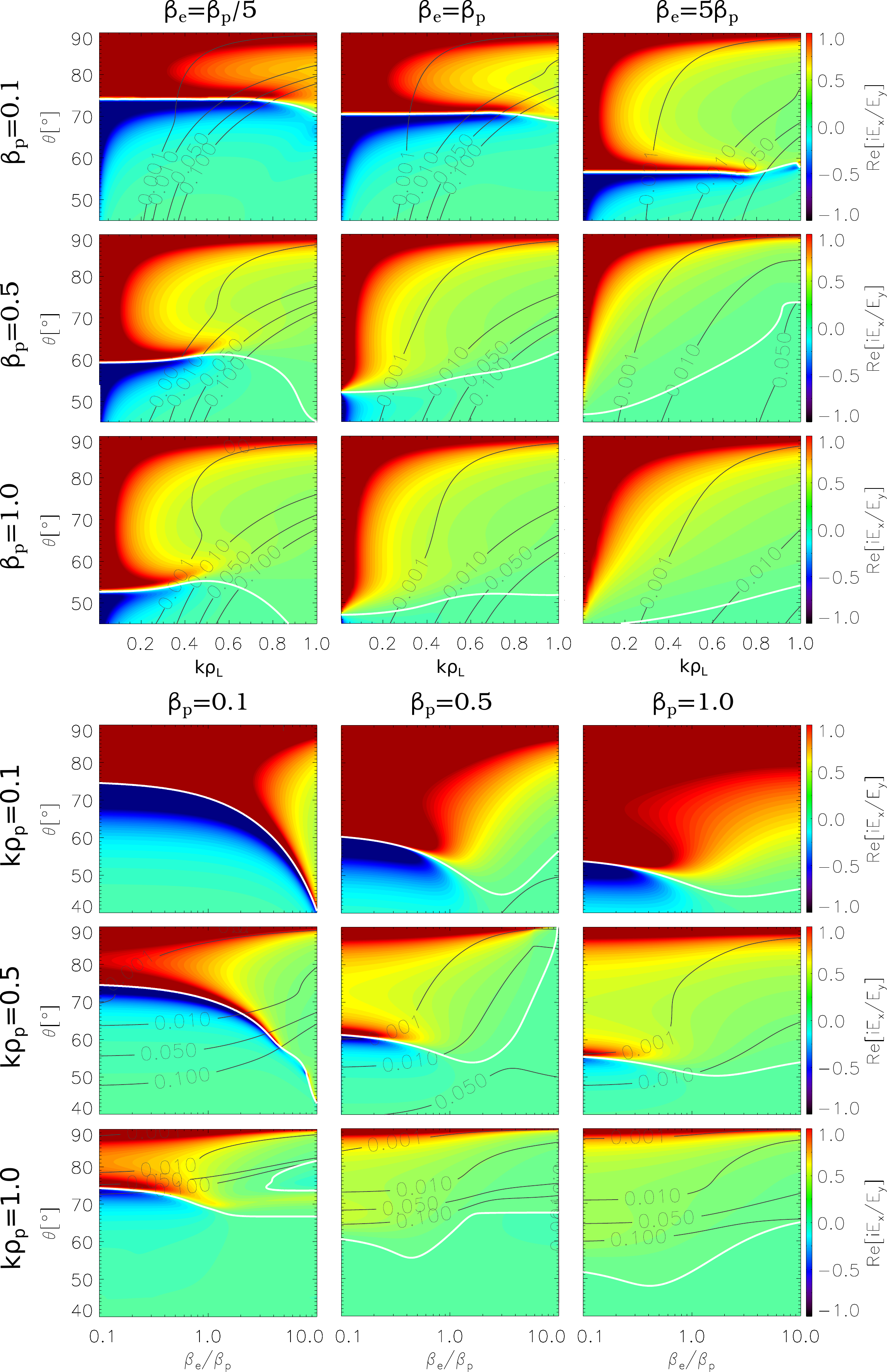}
\caption{Heat-maps displaying the transverse electric polarisation for a wide range of propagation angles and different combinations of $\beta_p$ and $\beta_e/\beta_p$ as a function of the wavenumber in units of $\rho_{L}^{-1}$ and for different combinations of $\beta_p$ and $k\rho_{p}$ as functions of $\beta_e/\beta_p$. The white lines indicate the contour of $\textnormal{Re}\left[iE_x/E_y\right]=0$ and black lines represent contours of characteristic values of $-\gamma/\Omega_p$, where $\gamma$ is the imaginary part of the wave frequency.}

\label{fig:maps_pol}
\end{figure*}

Figure \ref{fig:maps_pol} displays the mode's polarisation as heat maps as functions of the wavenumber, for fixed values of $\beta_p$, and of $T_e/T_p$, for fixed values of the wavenumber. The contours of $\textnormal{Re}\left[iE_x/E_y\right]=0$ are plotted in white, while the contours of characteristic values of the damping rate are overplotted in black.
The usual behavior of the transition from left-handed to right-handed polarisation of the Alfvénic waves in an electron-proton plasma, as seen in the early groundbreaking work by \cite{gary_1986} and a recent study by \cite{Villarroel_2023}, can be seen here for $\beta_p=0.1$. For this value of $\beta_p$, an increase in $\beta_e$ has the same effect as increasing $\beta$ for $T_e=T_p$. For higher values of $\beta_p$, the results exhibit a different behavior; at small wavenumbers, a rise in $T_e/T_p$ tends to bring the transition curve to smaller wave-normal angles, while for larger wavenumber, the transition occurs at larger propagation angles. This can be seen both in panels depicting the transverse electric polarisation as a function of the wavenumber and of the ratio $\beta_e/\beta_p$. It is also noteworthy that increasing the ratio $T_e/T_p$ tends to reduce the mode's damping rate in scales normalized to the ion-acoustic Larmor radius $\rho_L$, which depends on this ratio. When normalized to a value independent of $T_e$, however, an increase in the electron-to-proton temperature ratio will, in most cases, increase the wave's damping rates, as can be seen in Figures \ref{fig:disp} and \ref{fig:disp_fixedb}.

\begin{figure*}
\sidecaption
\includegraphics[width=12cm]{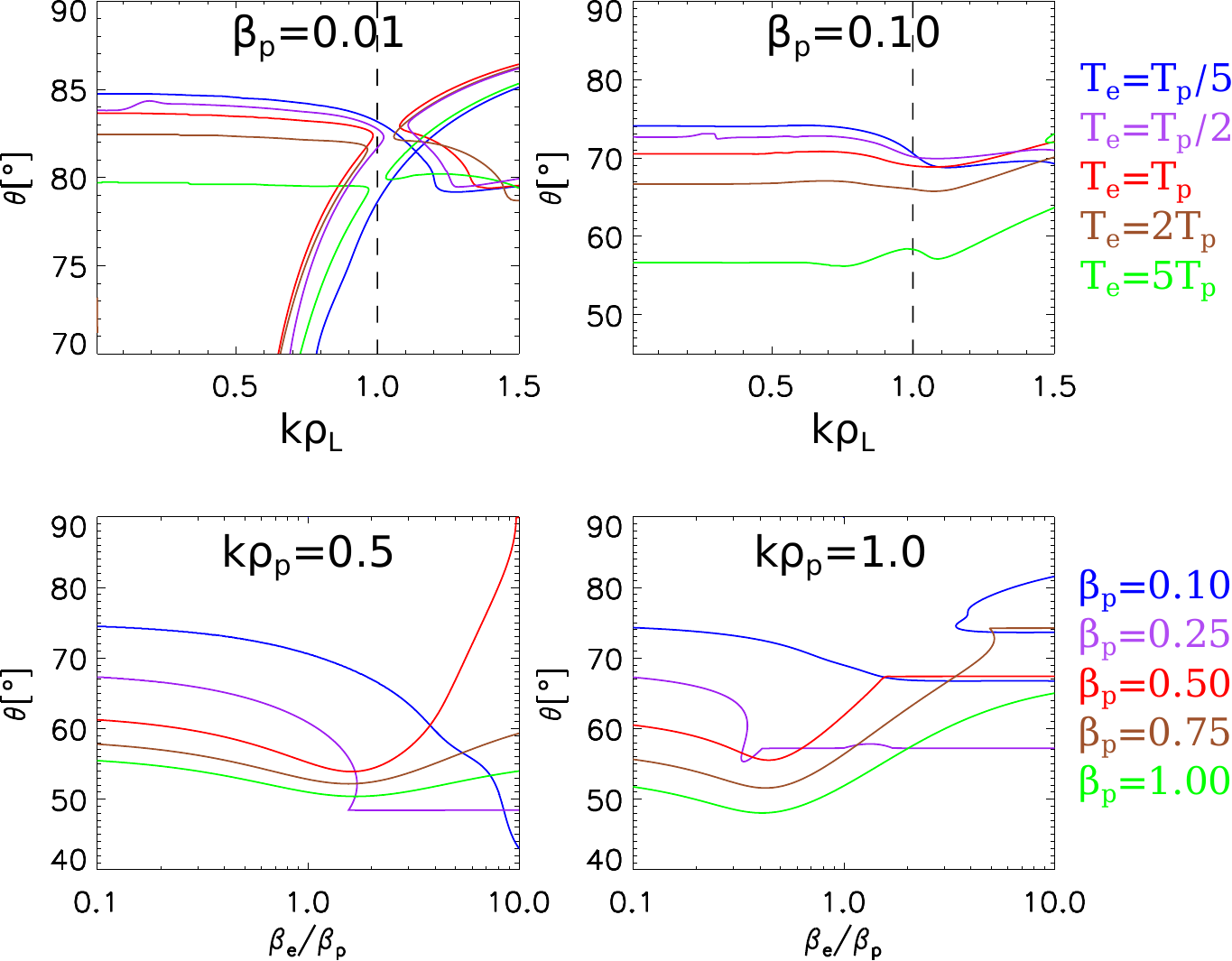}
\caption{Contours of $\textnormal{Re}\left[iE_x/E_y\right]=0$ for different electron-to-proton temperature ratios and $\beta_p$ as functions of the propagation angle and $k\rho_L$ in the top two panels and for different combinations of the $\beta_p$ and $k\rho_p$ as functions of the propagation angle and the ratio $\beta_e/\beta_p$ in the bottom two panels. }
\label{fig:maps_pol0}
\end{figure*}

Figure \ref{fig:maps_pol0} depicts the contours of $\textnormal{Re}\left[iE_x/E_y\right]=0$ with respect to the wavenumber and wave-normal angle for different combinations of $\beta_p$ and $T_e/T_p$ in the top row and with respect to $\beta_e/\beta_p$ and the wave-normal angle for different combinations of $ck/\omega_{pp}$ and $\beta_p$ in the bottom row.
For both $\beta_p=0.01$ and $\beta_p=0.10$, there is a segment of the transition curve for relatively large angles that is independent of the wavenumber until $k\rho_{L}\sim 1$, when kinetic effects dominate, and the transition curve becomes heavily dependant on the wavenumber in some complicated ways. This is not true for larger values of $\beta_p$, as can be deducted from Figure \ref{fig:maps_pol}, where the transition may depend on the wavenumber even in the large-scale domain of the waves. For larger values of $T_e/T_p$ and $\beta_p=0.10$, the transition tilts towards larger propagation angles. The same behavior can be appreciated for $\beta_p=0.01$, although in this case, the curve is disjointed, implying that there are certain wavenumbers around $\rho_L^{-1}$ for which $\textnormal{Re}\left[iE_x/E_y\right]>0$ and the transverse electric fields rotate in a right-handed sense at least for all wave-normal angles depicted in this Figure. We also observe that for $k\rho_L\geq1$ and larger wave-normal angles, the polarisation of the waves may shift from right-handed to left-handed, as can also be seen in Figure \ref{fig:disp}.
In the bottom panel, we see that an increase in $\beta_p$ reduces the propagation angle at which Alfvénic waves change from being left-hand to right-hand polarized for $\beta_p\geq\beta_e$. When $\beta_e>\beta_p$, the behavior tends to change, especially for larger values of $\beta_p$.

\section{Discussion} \label{sec:cite}
In this work, we present an extensive study on the effect of the thermal properties of electrons on the propagation of Alfvénic waves in warm, magnetized plasmas. Considering typical values of the electron-to-proton temperature ratio in space plasmas, such as Earth's magnetosphere and the solar wind, we show that the electron temperature can considerably modify the dispersive properties of KAW and EMIC waves, especially in the sub-protonic domain.

The results in Figures \ref{fig:disp} and \ref{fig:disp_fixedb} show that an increase in the electron-to-proton temperature ratio for fixed values of $\beta_p$ results in an increased wave damping with respect to the wave number. This is to be expected, as electron heating dominates the KAW's power dissipation \citep{Quataert_1998,Podesta_2012,Tong_2015, Zhao_2022}. The results also show that, for a fixed proton temperature, an increase in the electron temperature raises the frequency of the waves as the mode becomes dispersive, which is consistent with the previous \citep{Agarwal_2011}.
Comparing our results for the magnetic polarisation $iB_z/B_y$ to predictions from two-fluid theory in Figure \ref{fig:pol_bzy}, we show that an increase in the electron-to-proton temperature ratio allows the wave to acquire signature characteristics of the KAW at propagation angles significantly lower than $\theta\sim\pi/2$. For higher $\beta_p$ values, the results do not adjust well with the two-fluid theory predictions, although the difference between the three cases seems to be dependent on the combined $\beta$.  For the fluid prediction to adjust our results in the large beta case requires that the heat capacity ratios $\gamma_e,\gamma_p$ of electrons and protons, respectively, be both less than unity, which cannot represent any thermodynamic process in either a plasma or ideal gas. Thus, the discrepancy is attributed to a mismatch between fluid and kinetic theory in high-temperature plasmas. Thus, the discrepancy is assumed to be due to the limitations of fluid theories in the description of a high-temperature plasma. This is consistent with results by \cite{Krauss_1994}, which indicate that results from kinetic theory deviate from their fluid theory equivalents for $\beta\gtrsim 0.5$.
As seen on the different panels of Figure \ref{fig:disp}, the electron temperature plays an important role in the mode's transition from large scales, for which the solutions can be approximated by fluid theories, to the small scales that require a fully kinetic description. The mode becomes dispersive and wave damping appears at different wavenumbers for different electron-to-proton temperature ratios. In general, a larger electron temperature will cause the real frequency to depart from the MHD solution at smaller wavenumbers, and the same is true for the apparition of kinetic damping.  It is often mentioned in the literature that for the KAW, these effects arise when the perpendicular wavenumber becomes comparable to the inverse proton gyroradius or, equivalently, $k_\perp \rho_p\sim 1$. This criterion, however, completely excludes the thermal effects of electrons. Our results show that an increase in the electron temperature yields a similar outcome to that of raising $\beta$ in a plasma in which $T_p=T_e$. Thus, we conclude that $\beta_e$ contributes to a total $\beta=(\beta_e+\beta_p)/2$. A consequence of this is that a large value of $T_e/T_p$ allows the Alfvén mode to acquire signature characteristics of the KAW at smaller propagation angles than for $T_e/T_p\leq1$. The use of the ion-acoustic Larmor radius $\rho_L$ instead of $\rho_p$ provides an electron-temperature-sensible characteristic scale for the lower limit of the kinetic domain, which appears to be in good agreement with the departure of the kinetic solution from its MHD counterpart for the different temperature ratios considered, especially as the propagation angle becomes near-perpendicular and the mode can be identified as the KAW. This is further supported by the results displayed in Figure \ref{fig:disp_fixedb}. Here, the ion-acoustic $\beta$ is held fixed while the electron-to-proton temperature ratio varies. The results show only small differences in the real and imaginary frequencies, contrasting those of Figure \ref{fig:disp}. Also, as expected, the two-fluid prediction for the magnetic polarisation does not provide a faithful approximation to the kinetic solution for large $\beta$ values, as seen in the case where $\beta_e+\beta_p=1.20$. Interestingly, however, this does not hold true for the case with $T_e/T_p=5$, for which the two-fluid solution provides a good approximation at small wavenumbers. This is somewhat in contrast with \cite{Yang_Wu_Wang_Lee_2014}, which shows that fluid theory solutions agree better with kinetic theory results for $T_p\gg T_e$, although this study focuses on small $\beta$ values.
Figure \ref{fig:fields} provides useful insight into how the Aflvénic mode transitions from linear polarisation in MHD scales to elliptical polarisation in the kinetic domain. As expected from classical MHD, the mode's electric perturbations are sustained mostly by those in the wave's propagation direction in the perpendicular plane with respect to the background magnetic field ($E_x$, in this case). As the wave becomes dispersive, the mode acquires significant electric perturbations along the background field ($E_z$) and small but non-null perturbations orthogonal to the wave vector ($E_y$). This makes the polarisation elliptical, although the reason for the fields' rotations changing from left-hand to right-handed as the propagation angle increases is yet unclear. With the larger $\beta$ values, the $E_y$ field becomes larger, even becoming the largest component of the electric perturbations for some of the studied configurations. This makes the $E_x$ field decrease in magnitude, thus making the transverse electric polarisation of the waves become almost linear for large wavenumbers and even change its sign from right-hand to left-handed, as can be seen for large $T_e/T_p$ in Figure \ref{fig:disp}. It is also worth noting that the parallel electric perturbations arise even for smaller propagation angles than those that can be considered as propitious for KAW propagation, but for relatively small $\beta$ values, only near perpendicular propagation does the magnitude of this electric component have maxima in sub-protonic scales.
Figure \ref{fig:beta_pol} displays relevant quantities in the study of the KAW as functions of the electron-to-proton temperature ratios. These results imply that high $T_e/T_p$ tends to increase the value of the field-aligned electric perturbations. We also note that varying $T_e/T_p$ at lower wave-normal angles may reverse the sign of the transverse electric polarisation. It is noteworthy, however, that the nature of the changes in the rotation sense of the transverse electric fields is different for $\beta_p=0.1$ and $\beta_p=0.5$. In the former case, the behavior in the polarisation suggests that the $E_y$ component becomes zero, something which is supported by the fact that $i\frac{B_z}{B_y}$, which has $E_y$ in its numerator through Faraday's law, becomes zero at the same electron-to-proton temperature ratio. \\
Figure \ref{fig:maps_pol} displays the transverse electric polarisation as heat maps, with contours of $\textnormal{Re}\left[iE_x/E_y\right]=0$ in white, indicating the transition from left-hand to right-hand polarized Alfvén waves and damping rate contours in black. As expected, for small $\beta$ values, an increase in $T_e/T_p$ has the same effect as increasing $\beta$ in an isothermic plasma, as reported in \cite{Villarroel_2023}. For larger betas, the result tends to be counter-intuitive; for smaller wavenumbers, augmenting $T_e/T_p$ lowers the transition curve, while for larger wavenumbers, this curve drifts towards larger propagation angles. This is corroborated by the results for fixed wavenumbers as functions of $\beta_e/\beta_p$. It is also interesting to note that, although under the same normalization, an increase in $T_e/T_p$ results in stronger Landau damping (Figures \ref{fig:disp}, \ref{fig:disp_fixedb}) when normalized to $\rho_L$, the opposite holds.\\
Figure \ref{fig:maps_pol0} exhibits different $\textnormal{Re}\left[iE_x/E_y\right]=0$ contours for different plasma configurations, for fixed $\beta_p$ with respect to the propagation angle and wavenumber and for fixed wavenumber with respect to the propagation angle and electron-to-proton temperature ratio. The results show that normalizing to $\rho_L$ allows us to identify a region where the transition from left-hand to right-hand polarized Alfvén waves is independent of the wavenumber. For small values of $\beta$, there are two disjointed regions where left-handed polarisation is allowed, which implies that there are particular wavenumbers near $\rho_L^{-1}$ for which the waves acquire right-handed polarisation for a wide range of propagation angles. 
From the bottom panels of this Figure, we observe that an increase in $T_e/T_p$ has a similar effect to raising the total $\beta$ only until $T_e\sim T_p$ for high values of $\beta_p$. Kinetic effects related to electrons are likely the reason why this does not hold for $T_e>T_p$ in cases where $\beta\gtrsim 0.5$. \\
In summary, we have shown that the thermal properties of electrons have strong and non-trivial effects on the dispersive properties of Alfvénic waves in the kinetic domain. We propose that electrons contribute to a total $\beta$ as in two-fluid theories, which sets the gateway of the kinetic domain through an ion-acoustic Larmor radius $\rho_L$. Fixing the total $\beta$ results in very similar dispersion relations with different electron-to-proton temperature ratios, while only fixing $\beta_p$ results in very different ones. The results from kinetic theory deviate from the fluid picture at large $\beta$ values, which is consistent with previous results.

\begin{acknowledgements}
We thank the support of ANID, Chile, through National Doctoral Scholarship N° 21220616 (NVS) and Fondecyt Initiation Grant 11201048 (RAL), and to the Research Vice-Rectory of the University of Chile (VID) through Grant ENL08/23 (PSM).
\end{acknowledgements}

%

\vspace{5mm}
\bibliographystyle{aa}
\bibliography{bibliography}

\end{document}